\documentclass[a4paper]{article}
\usepackage{amsmath,graphicx,multirow,booktabs,float,threeparttable}
\newcommand{\tabincell}[2]{\begin{tabular}{@{}#1@{}}#2\end{tabular}}
\usepackage{lipsum}

\usepackage{INTERSPEECH2019}

   \title{The DKU System for the Speaker Recognition Task of \\ the 2019 VOiCES from a Distance Challenge}
    \name{Danwei Cai$^1$, Xiaoyi Qin$^{1, 2}$, Weicheng Cai$^{1, 2}$, Ming Li$^1$\thanks{This research was funded in part by the National Natural Science Foundation of China (61773413), Natural Science Foundation of Guangzhou City (201707010363), Six talent peaks project in Jiangsu Province (JY-074), Science and Technology Program of Guangzhou City (201903010040), and Huawei. We also thank Weixiang Hu, Yu Lu, Zexin Liu and Lei Miao from Huawei Digital Technologies Co., Ltd, China.}
}
\address{
  $^1$Data Science Research Center, Duke Kunshan University, Kunshan, China\\
  $^2$School of Electronics and Information Technology, Sun Yat-sen University, Guangzhou, China }
\email{ming.li369@dukekunshan.edu.cn}

\begin{document}

\maketitle

\begin{abstract}
In this paper, we present the DKU system for the speaker recognition task of the VOiCES from a distance challenge 2019. We investigate the whole system pipeline for the far-field speaker verification, including data pre-processing, short-term spectral feature representation, utterance-level speaker modeling, back-end scoring, and score normalization. Our best single system employs a residual neural network trained with angular softmax loss. Also, the weighted prediction error algorithms can further improve performance. It achieves 0.3668 minDCF and 5.58\% EER on the evaluation set by using a simple cosine similarity scoring. Finally, the submitted primary system obtains 0.3532 minDCF and 4.96\% EER on the evaluation set.
\end{abstract}

\section{Introduction}

In the past decade, the performance of speaker recognition has improved significantly. The i-vector based method \cite{dehak_front-end_2011} and the deep neural network (DNN) based methods \cite{snyder_x-vectors:_2018, cai_exploring_2018} have promoted the development of speaker recognition technology in telephone channel and closed talking scenarios. However, speaker recognition under far-field and complex environmental settings is still challenging due to the effects of the long-range fading, room reverberation, and complex environmental noises. Speech signal propagating in long-range suffers from fading, absorption, and reflection by various objects, which change the pressure level at different frequencies and degrade the signal quality \cite{woelfel_distant_2009}. Reverberation includes eaarlay reverberation and late reverberation. Early reverberation (i.e., reflections within 50 to 100 ms after the direct wave arrives at the microphone) can improve the received speech quality, while late reverberation will degrade the speech quality. The adverse effects of reverberation on speech signal includes smearing spectro-temporal structures, amplifying the low-frequency energy, and flattening the formant transitions, etc. \cite{assmann_perception_2004}. Also, the complex environmental noises ``fill in'' regions with low speech energy in the time-frequency plane and blur the spectral details \cite{woelfel_distant_2009}. These effects result in the loss of speech intelligibility and speech quality, imposing great challenges in far-field speaker recognition and far-field speech recognition.

To compensate for the adverse impacts of room reverberation and environmental noise, various approaches have been proposed at different stages of the speaker recognition system. At the signal level, dereverberation \cite{WPE}, denoising \cite{zhao_robust_2014, kolboek_speech_2016, oo_dnn-based_2016, eskimez_front-end_2018}, and beamforming \cite{heymann_neural_2016, warsitz_blind_2007} can be used for speech enhancement. At feature level, sub-band Hilbert envelopes based features~\cite{falk_modulation_2010, sadjadi_hilbert_2011}, warped minimum variance distortionless response (MVDR) cepstral coefficients \cite{jin_speaker_2010}, blind spectral weighting (BSW) based features \cite{sadjadi_blind_2014} have been applied to ASV system to suppress the adverse impacts of reverberation and noise. At the model level, reverberation matching with multi-condition training models has been successfully employed within the universal background model (UBM) or i-vector based front-end systems \cite{peer_reverberation_2008, avila_improving_2014}. In back-end modeling, multi-condition training of probabilistic linear discriminant analysis (PLDA) models were employed in i-vector system~\cite{garcia-romero_multicondition_2012}. The robustness of deep speaker embeddings for far-field speech has also been investigated in \cite{nandwana_robust_2018}. Finally, at the score level, score normalization \cite{peer_reverberation_2008} and multi-channel score fusion \cite{jin_far-field_2007, mikyong_ji_text-independent_2008} have been applied in far-field ASV system to improve the robustness.

The ``VOiCES from a Distance Challenge 2019'' is designed to foster research in the area of speaker recognition and automatic speech recognition (ASR) with the special focus on single channel far-ﬁeld audio, under noisy conditions \cite{nandwana_voices_2019}. Our system pipeline consists of the following six main components, including data pre-processing, short-term spectral feature extraction, utterance-level speaker modeling, back-end scoring, score normalization, as well as fusion and calibration.

 This paper is organized as follows: Section 2 describes the details of our submitted system. Section 3 clarifies the data usage, with experimental results and analysis. Conclusions are drawn in section 4.

\section{System descriptions}

\subsection{Data pre-processing}

\subsubsection{Data augmentation}
We adopt two kinds of data augmentation strategies. The first is the same as the x-vector system available at Kaldi Voxceleb recipe, which employs additive noises and reverberation.
We also use \textit{pyroomacoustics} \cite{pyroomacoustics} to simulate the room acoustic based on RIR generator using Image Source Model (ISM) algorithm. The microphones, distractors, and speech source are similar to the room settings presented in \cite{richey_voices_2018}. We use the music and noise part of the MUSAN dataset \cite{musan} to generate the television noise, and the `us-gov' part to create babble noise.

For the systems described below, we use the Kaldi data augmentation strategy for the MFCC i-vector system and the TDNN x-vector system, and \textit{pyroomacoustics} data augmentation strategy for the remaining systems.

\subsubsection{Dereverberation}
The weighted prediction error (WPE) algorithm is a successful algorithm to reduce late reverberation \cite{WPE}. The method estimates the optimal dereverberation filter coefficients based on iterative optimization. During the enrolling and testing, we use the single-channel WPE to dereverberate the sound with a dereverberation filter of 10 coefficients. The WPE codes are from \texttt{http://www.kecl.ntt.co.jp/icl/signal/wpe}.

\subsection{Short-term spectral feature}
 Four features including Mel-frequency cepstral coefficient~(MFCC), power-normalized cepstral coefficients~(PNCC), Mel-filterbank energies (Mfbank) and gammatone-Filterbank energies (Gfbank) are adopted in our systems.
\subsubsection{MFCC}
Two kinds of MFCC features with a different number of cepstral filterbanks are adopted, which result in 20- and 30-dimensional MFCCs (MFCC-20 and MFCC-30). MFCC-20 is for the i-vector system, and MFCC-30 is for the TDNN x-vector system. Short-time cepstral mean subtraction~(CMS) over a 3-second sliding window is applied. For the MFCC-20, their first and second derivatives are computed before applying the CMS.

\subsubsection{PNCC}
PNCC has proved to be more robust in various types of additive noise and reverberant environments compared to MFCC in ASR~\cite{PNCC}. The major features of PNCC processing include the use of a power-law nonlinearity that replaces the traditional log nonlinearity used in MFCC coefficients, a noise-suppression algorithm based on asymmetric filtering that suppress background excitation, and a module that accomplishes temporal masking~\cite{PNCC}. 20-dimensional PNCC are extracted using a 25 ms window with 10 ms shifts. First and second derivatives are computed before applying CMS.

\subsubsection{Log Mel-filterbank energies}
Each audio is converted to 64-dimensional log Mel-filterbank energies with cepstral filterbanks ranging from 20 to 7600 Hz (Mfbank-16k). We also downsample the audio to 8000 sample rate and use cepstral filter banks within the range of 20 to 3800 Hz to calculate Mfbank-8k features. A short-time cepstral mean subtraction is applied over a 3-second sliding window.

\subsubsection{Gammatone-Filterbank Energies}
Gammatone filters are approximations to the filtering system of human ear \cite{patterson_complex_1992}. The Gammatone filterbanks are selected within the range of 50 to 8000 Hz to compute the 64-dimensional Gammatone-filterbank energies. Short-time CMS is then applied over a 3-second sliding window.

\subsection{Utterance-level speaker modeling}
We extract the utterance-level speaker embeddings from three state-of-the-art modelings, including the i-vector system~\cite{dehak_front-end_2011},  the TDNN x-vector system~\cite{snyder_x-vectors:_2018}, and the deep ResNet system~\cite{cai_exploring_2018}. 

\subsubsection{i-vector}
We train two i-vector systems on the MFCC-20 and PNCC features respectively. The extracted 60-dimensional features are used to train a 2048 component Gaussian mixture model-universal background model (GMM-UBM) with full covariance matrices. Then zero-order and first-order Baum-Welch statistics are computed on the UBM for each recording's MFCC feature, and single factor analysis is employed to extract i-vectors with 600 dimensions \cite{dehak_front-end_2011}.

\subsubsection{TDNN x-vector}
The x-vector system is developed by adapting the Kaldi Voxceleb recipe. For the x-vector extractor, a DNN is trained to discriminate speakers in the training set. The first five timed delayed layers operate at frame-level. Then a temporal statistics pooling layer is employed to compute the mean and standard deviation over all frames for an input segment. The resulted segment-level representation is then fed into two fully connected layers to classify the speakers in the training set. After training, speaker embeddings are extracted from the 512-dimensional affine component of the first fully connected layer.

\begin{table*}[t]
    \caption{Development subset results for the speaker recognition task of the VOiCES from a distance challenge (SN represents Score Normalization, devW represents whitening using development subset)}
    \label{tab: results1}
    \centering
    \begin{threeparttable}
    \begin{tabular}[c]{@{\ \ }cl@{\ \ }cc ccc ccc@{\ \ }}
        \toprule
        \multirow{3}*{\textbf{Front-end}} & \multirow{3}*{\textbf{Back-end}} & \multirow{3}*{\textbf{WPE}} & \multirow{3}*{\textbf{SN}} & \multicolumn{3}{c}{\textbf{Development subset}} & \multicolumn{3}{c}{\textbf{Evaluation}} \cr
        \cmidrule(lr){5-7} \cmidrule(l){8-10}
        & & & & \textbf{minC} & \textbf{actC} & \textbf{EER[\%]} & \textbf{minC} & \textbf{actC} & \textbf{EER[\%]} \\
        \midrule  
        \multirow{2}*{\tabincell{c}{MFCC i-vector}} & PLDA & - & $\surd$ & 0.4935 & 0.6747 & 6.33 & 0.8037 & 0.8294 & 12.92\\
         & CORAL + devW + PLDA & $\surd$ & $\surd$ & 0.4527 & 0.4703 & 6.12 & 0.6870 & 0.6891 & 11.89 \\
        \midrule 
        \multirow{2}*{\tabincell{c}{PNCC i-vector}} & PLDA & - & $\surd$ & 0.5073 & 0.6745 & 6.12 & 0.6791 & 0.7803 & 10.18 \\
         & CORAL + devW + PLDA & $\surd$ & - & 0.4594 & 0.4697 & 5.29 & 0.6498 & 0.7152 & 10.09 \\
        \midrule 
        \multirow{2}*{x-vector} & CORAL + PLDA & - & $\surd$ & 0.4018 & 0.4151 & 4.96 & 0.6377     & 0.6492 & 09.13 \\
         & CORAL + PLDA & $\surd$ & - & 0.3617 & 0.3688 & 4.52 & 0.5417 & 0.5544 & 07.54 \\
        \midrule 
        \multirow{2}*{\tabincell{c}{Mfbank-8k \\ ResNet + Softmax}} & CORAL + devW + PLDA & - & - & 0.4557 & 0.5246 & 5.41 & 0.6608 & 0.7128 & 10.92 \\
         & CORAL + devW + PLDA & $\surd$  & - & 0.3934 & 0.4611 & 4.59 & 0.5929     & 0.6424 & 09.75 \\
        \midrule 
        \multirow{2}*{\tabincell{c}{Mfbank-16k \\ ResNet + Softmax}} & cosine similarity & - & - & 0.3608 & 1 & 3.81 & 0.6262 & 1 & 08.75 \\
         & cosine similarity & $\surd$ & - & 0.3245 & 1 & 3.02 & 0.5507     & 1 & 07.91 \\
        \midrule 
        \multirow{2}*{\tabincell{c}{Mfbank-16k \\ ResNet + A-Softmax}} & cosine similarity & - & - & 0.2735 & 1 & \textbf{2.73} & 0.4156 & 1 & \textbf{05.84} \\
         & cosine similarity & $\surd$ & - & \textbf{0.2485} & 1 & \textbf{2.41} & \textbf{0.3668} & 1 & \textbf{05.58} \\
        \midrule 
        \multirow{2}*{\tabincell{c}{Gfbank \\ ResNet + A-Softmax}} & cosine similarity & - & - & 0.3065 & 1 & 3.52 & 0.4411 & 1 & 06.78    \\
         & cosine similarity & $\surd$ & - & \textbf{0.2680} & 1 & 3.14 & \textbf{0.4056} & 1 & 06.49 \\
        \bottomrule
    \end{tabular}
    \end{threeparttable}
\end{table*}

\subsubsection{Deep ResNet}

We follow the deep ResNet system as described in~\cite{cai_insights_2018,cai_exploring_2018,Cai_2018_Interspeech}, and we increase the widths (number of channels) of the residual blocks from \{16, 32, 64, 128\} to \{32, 64, 128, 256\}.  The network architecture contains three main components: a front-end ResNet, a pooling layer, and a feed-forward network. The front-end ResNet transforms the raw feature into a high-level abstract representation. The subsequent pooling layer outputs a single utterance-level representation. Specifically, means statistics are accumulated for each feature map, and finally 256-dimensional utterance-level representation is produced. Each unit in the output layer is represented as a target speaker identity. 

All the components in the pipeline are jointly learned in an end-to-end manner with a unified loss function. We adopt the typical softmax loss as well as the angular softmax loss (A-softmax) \cite{liu_sphereface:_2017}. A-softmax learns angularly discriminative features by generating an angular classiﬁcation margin between embeddings of different classes. The superiority of A-softmax has been shown in both face recognition \cite{liu_sphereface:_2017}, language recognition and speaker recognition \cite{cai_exploring_2018}.

After training, the 256-dimensional utterance-level speaker embedding is extracted after the penultimate layer of the neural network for the given utterance. In the testing stage, the full-length feature sequence is directly fed into the network, without any truncate or padding operation.

Based on the deep ResNet framework, we investigate multiple kinds of short-term spectral features and loss functions. Finally, we have four networks trained with different setups: 

\begin{itemize}
\item Mfbank-8k + Softmax: ResNet system trained on Mfbank-8k features with softmax loss.
\item Mfbank-16k + Softmax: ResNet system trained on Mfbank-16k features with softmax loss.
\item Mfbank-16k + A-softmax: ResNet system trained on Mfbank-16k features with A-softmax loss.
\item Gfbank + A-softmax. ResNet system trained on Gfbank- features with A-softmax loss.
\end{itemize}

\subsection{Back-end modeling}
In back-end modeling, we either use cosine similarity based scoring, or Probabilistic Linear Discriminant Analysis (PLDA) based scoring.

\subsubsection{Cosine similarity}
We use cosine similarity as a scoring method for the ResNet based systems. The scores of any given enrollment-test pair are calculated as the cosine similarity of the two embeddings.

\subsubsection{Gaussian PLDA}
We use Correlation Alignment (CORAL) \cite{sun_return_2016, alam_speaker_2018} to align the distributions of out-of-domain and in-domain features in an unsupervised way by aligning second-order statistics, i.e., covariance. To minimize the distance between the covariance of the out-of-domain and in-domain features, a linear transformation $\mathbf{A}$ to the original source features and the Frobenius norm is used as matrix distance metric:
\begin{equation}
\min_\mathbf{A}\| \mathbf{C}_{\hat{S}} - \mathbf{C}_{T} \|_{F}^{2} = \min_\mathbf{A}\| \mathbf{A}^T\mathbf{C}_{S}\mathbf{A} - \mathbf{C}_{T} \|_{F}^{2}
\end{equation}
where $\mathbf{C}_{S}$ and $\mathbf{C}_{T}$ are covariance matrix of the source-domain and target-domain features, $\mathbf{C}_{\hat{S}}$ is covariance of the transformed source features, and $\|\cdot\|_{F}^{2}$ denotes the matrix Frobenius norm.

The embeddings after domain adaptation are whitened and unit-length normalized. The whitening transforms is estimated with either the training set or the development subset. 

The Gaussian PLDA model \cite{garcia-romero_analysis_2011} with a full covariance residual noise term is trained on the speaker discriminant features. After the PLDA is trained, the scores of any given enrollment-test pair are calculated as the log-likelihood ratio on the PLDA model.

\subsection{Score normalization}
After scoring, results from all trials are subject to score normalization. We utilize Adaptive Symmetric Score Normalization (AS-Norm) in our systems \cite{sn_analysis_2017}. The adaptive cohort for the enrollment file are selected to be $X$ closest (most positive scores) files to the enrollment utterance $e$ as $\mathcal{E}_{e}^{\text{top}}$. The cohort scores based on such selections for the enrollment utterance are then:
\begin{equation}
    S_e(\mathcal{E}_{e}^{\text{top}}) = \{s(e, \varepsilon)  | \forall \varepsilon \in \mathcal{E}_{e}^{top}\}
\end{equation}

Then the AS-Norm is
\begin{equation}
\begin{aligned}
    \tilde{s}(e, t) = & \frac{1}{2} \left( \frac{s(e, t) - \mu[S_e(\mathcal{E}_{e}^{\text{top}})] }{ \sigma[S_e(\mathcal{E}_{e}^{\text{top}})] } + \frac{s(e, t) - \mu[S_t(\mathcal{E}_{t}^{\text{top}})] }{ \sigma[S_t(\mathcal{E}_{t}^{\text{top}})] }\right)
\end{aligned}
\end{equation}

\subsection{System fusion and calibration}
All the subsystems are fused and calibrated using the BOSARIS toolkit \cite{brummer2013bosaris} which learn a scale and a bias for each subsystem. The final fusion is a score-level equal-weighted sum after applying the scale and the bias.

\begin{figure*}[t]
  \centering
  \makebox[\textwidth][c]{\includegraphics[width=530pt]{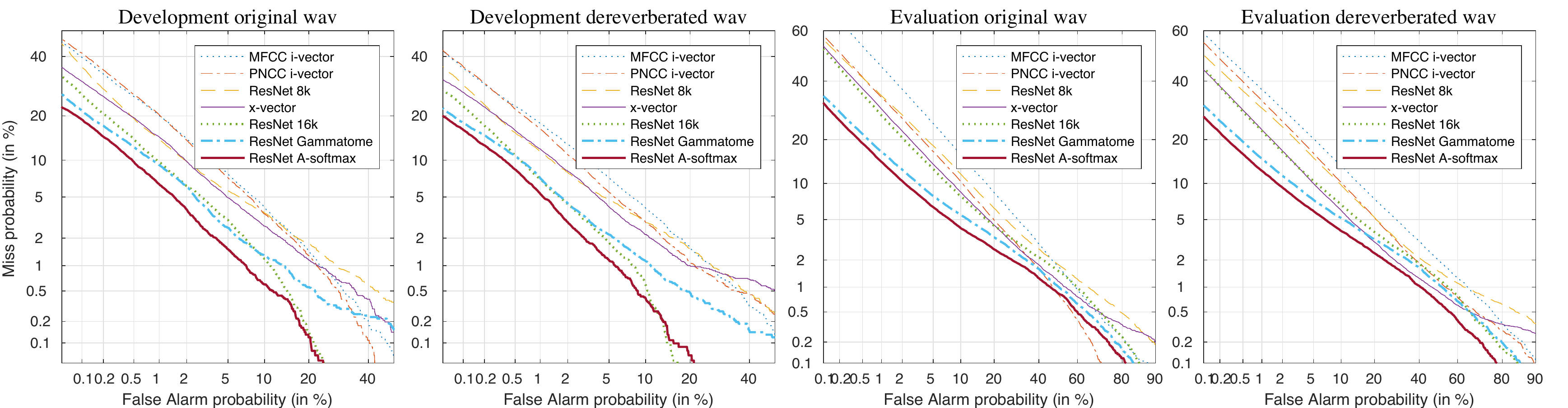}}
  \caption{DET plots for development and evaluation dataset with original or dereverberated sound wave}
  \label{fig: det}
\end{figure*}

\begin{table*}[t]
  \caption{System performance on different fusion system}
  \centering
  \label{tab: results2}    
    \begin{threeparttable}
    \begin{tabular}[c]{lcc@{\ \ }c@{\ \ }c  cc@{\ }c@{\ \ }c}
        \toprule
        \multirow{3}*{\textbf{Fusion strategy}} & \multicolumn{4}{c}{\textbf{Development subset}} & \multicolumn{4}{c}{\textbf{Evaluation}} \cr
        \cmidrule(lr){2-5} \cmidrule(lr){6-9}
        & \textbf{minC} & \textbf{actC} & \textbf{EER[\%]} & \textbf{Cllr} & \textbf{minC} & \textbf{actC} & \textbf{EER[\%]} & \textbf{Cllr} \\
        \midrule
        Best single system (ResNet + A-softmax + WPE) & 0.2485 & 1 & 2.41 & 0.8060 & 0.3668 & 1 & 5.58 & 0.8284 \\
        \midrule
        Each embedding with top 1 back-end & 0.1831 & 0.1857 & 1.93 & 0.0808 & \textbf{0.3205} & \textbf{0.3214} & \textbf{4.60} & \textbf{0.2335} \\
        Each embedding with top 2 back-end & 0.1644 & 0.1659 & 1.48 & 0.0710 & 0.3555 & 0.3578 & 4.79 & 0.2684 \\
        Each embedding with top 3 back-end (submission) & \textbf{0.1473} & \textbf{0.1484} & \textbf{1.21} & \textbf{0.0577} & 0.3532 & 0.3609 & 4.96 & 0.2683 \\
        \bottomrule
    \end{tabular}
    \end{threeparttable}
\end{table*}

\section{Experiments}
\subsection{Data usage}

The training data includes VoxCeleb 1~\cite{nagrani_voxceleb:_2017} and VoxCeleb 2~\cite{chung_voxceleb2:_2018}. The original distribution of VoxCeleb split each video into multiple short segments. During training, the segments from the same video are concatenated into a single sound wave, which results in 167897 utterances from 7245 speakers. No voice activity detection (VAD) is applied.

For the development data, we only use a subset of the development dataset provided by the VOiCES challenge. The total of 196 speakers in the original development dataset is split into two subgroups, each with 98 speakers. One subset is used as the new development set, and the other is used as the domain adaptation and score normalization corpus. In this way, we reduce the original 4,005,888 trials into 999,424 trials. Since a part of the development, data is used as the domain adaption and score normalization data, we can not provide the experimental results on the whole development data. So \emph{all the experimental results on the development set presented in this paper use the new sub-trials.}

\subsection{System performance on single systems}
In table \ref{tab: results1}, the systems of different front-end speaker discriminant features with the top one back-end are provided.

From the results in table \ref{tab: results1}, several observations are drawn as follows. First, the PNCC based i-vector system obtains a noticeable performance gain under strong reverberation and low SNR (signal to noise ratio) environments (evaluation set) compared to MFCC based i-vector system. For the development set with mild reverberation and higher SNR (about 20dB), the performance gain is not so obvious. Also, the WPE dereverberation algorithm results in 10\% gain compared to the original wave for both i-vector and neural network based systems. Moreover, the ResNet + softmax system trained on 16k Mfbank achieves 17.5\% relative performance gain in terms of minDCF compared to the 8k Mfbank. The last observation from the results is the performance of the system with A-softmax loss. Compared to the ResNet + softmax system, the ResNet + A-softmax system significantly improve the system performance by more than 20\% on both development and evaluation sets.

 The Detection Error Tradeoff (DET) curves in figure \ref{fig: det} provide a clear comparison among the subsystems we used in the VOiCES challenge.

The final best signal system is the ResNet + A-softmax network combined with cosine similarity scoring. Applying dereverberation to the enrollment and testing data can further improve the performance. On the development set, the final minDCF and EER are 0.2485 and 2.41\% respectively. On the evaluation set, the final minDCF and EER are 0.3668 and 5.58\%.

The performance degradation on the evaluation set can be observed from results. This performance degradation mainly due to the more challenging reverberation environments and much lower SNR in the evaluation data, which lead to the mismatch between development and evaluation data.

\subsection{System performance on fused systems}
For the seven kinds of front-end systems, the embeddings from the original audio and the de-reverberated audio are extracted respectively, resulting in 14 types of front-end speaker discriminant features. Then, different back-end modeling methods, including cosine scoring, a different set of PLDA modeling, and different setting of score normalization, are applied to these features. For each speaker embedding, the top three back-end methods with the best performance on the particular embedding are selected, and finally, we get 42 individual scores for the final fusion.

The final results on the development subset and the evaluation set are shown in table \ref{tab: results2}. Our final submission obtains minDCF of 0.1473 and 0.3532 on the development and evaluation set respectively.
	
After the evaluation, we investigate the system performance fused with different back-ends. It is interesting to find that although fusion with the top 3 back-ends for each front-end embeddings improves the performance by 20\% relatively compared to fusion with top 1 back-ends, the results on the evaluation show the opposite: fusion with the top 3 back-ends for each front-ends degrades the performance by 10\% compared to the fused system with top 1 back-ends. This is mainly because of the mismatch between the development and evaluation data.

\section{Conclusions}
We presented the components and analyzed the results of the DKU-SMIIP speaker recognition system for the VOiCES from a Distance Challenge 2019. We use different acoustic features, different front-end modeling methods, and various back-end scoring methods. To further improve the performance, we use WPE to dereverberate the development and evaluation data. This enabled a series of incremental improvements, and the fusion showed that different subsystems are complementary to each other at score level.

\bibliographystyle{IEEEtran}
\bibliography{mybib}

\end{document}